\documentclass[twocolumn]{aastex631}

\usepackage{enumitem}
\usepackage{graphicx}
\usepackage{amsmath}
\usepackage{comment}
\usepackage{txfonts}
\usepackage{booktabs}

\usepackage{natbib}
\usepackage{xcolor}
\usepackage{subcaption}
\usepackage{caption}
\usepackage{url}

\bibpunct{(}{)}{;}{a}{}{,}

\shorttitle{High-pass filter periodogram}

\begin{document}

\title{High-pass filter periodogram: an improved power spectral density estimator for unevenly sampled data.}


\author[0000-0002-7816-6401]{Ezequiel Albentosa-Ruiz}
\affiliation{Dpt. Astronomia i Astrof\'isica, Universitat de Val\`encia, C/ Dr. Moliner 50, Valencia, Spain}

\author[0000-0002-5523-7588]{Nicola Marchili}
\affiliation{Italian ALMA Regional Centre, INAF-Istituto di Radioastronomia, Via P. Gobetti 101, I-40129 Bologna, Italy}

\begin{abstract}
Accurate time series analysis is essential for studying variable astronomical sources, where detecting periodicities and characterizing power spectral density (PSD) are crucial. The Lomb-Scargle periodogram, commonly used in astronomy for analyzing unevenly sampled time series data, often suffers from noise introduced by irregular sampling. This paper presents a new high-pass filter (HPF) periodogram, a novel implementation designed to mitigate this sampling-induced noise. By applying a frequency-dependent high-pass filter before computing the periodogram, the HPF method enhances the precision of PSD estimates and periodicity detection across a wide range of signal characteristics. Simulations and comparisons with the Lomb-Scargle periodogram demonstrate that the HPF periodogram improves accuracy and reliability under challenging sampling conditions, making it a valuable complementary tool for more robust time series analysis in astronomy and other fields dealing with unevenly sampled data.
\end{abstract}


\section{Introduction}
\label{sec:intro}

Time series analysis plays an essential role in the study of variable astronomical sources.

The detection of possible periodicities and the correct characterisation of the power spectral density (PSD) of the signal emitted by a source are among the most valuable pieces of information that could be gained through time series analysis; they can provide us with essential indications concerning the physical processes occurring in the emitting regions, and help us to discriminate between different emission models. This explains the crucial position occupied in variability studies by the periodogram (see \citealt{1898TeMag...3...13S} for the classical expression; for uneven data, it is mostly used in the formulation known as Lomb-Scargle periodogram, see \citealt{1976Ap&SS..39..447L} and \citealt{ScargleII1982}), which is able to extract from the light curves information about both periodicities and PSD slope.

Similarly to any other tool of time series analysis, the periodogram is affected by the sampling and the length of the analysed datasets. The finite duration of the observations gives rise to sidelobes, i.e. a spectral leakage to nearby frequencies. The most serious issue for a reliable characterisation of the variability, however, is caused by the sampling. \cite{2018ApJS..236...16V} provides a clear and thorough discussion of how the sampling affects the spectral information returned by Discrete Fourier Transform, classical periodogram, and Lomb-Scargle periodogram. When the sampling of the analysed datasets is even, the spectral power is transferred to distant frequencies through aliasing; the frequencies to which the power is transferred are strictly related to the sampling interval $\Delta$t, as they are multiples of 1/$\Delta$t. In the case of uneven sampling, the power transfer caused by the sampling is randomised by the different time intervals between consecutive observations, leading to non-structured frequency peaks whose superposition results in a large and unpredictable increase of noise at high frequencies; this can completely hide the real characteristics of the underlying signal.

The development of the Lomb-Scargle periodogram proceeds from the need to find a way to retrieve reliable spectral information from astronomical datasets, whose sampling is generally uneven. The main purpose of the modification proposed by \cite{ScargleII1982}, and based on the previous work of \cite{1976Ap&SS..39..447L}, is to make sure that the simple statistical behaviour of the evenly spaced case remains valid also with uneven samplings. The Lomb-Scargle (LS) periodogram (from now on, we will refer to it simply as "the periodogram"; the classical algorithm by \citealt{1898TeMag...3...13S} will not be further considered) at angular frequency $\omega$ is defined as follows:

\begin{align}
    P_X (\omega) = \frac{1}{2} \Bigg\{ & \frac{ \left[ \sum_j X_j \cos \omega (t_j - \tau) \right]^2 } { \sum_j \cos^2 \omega (t_j - \tau) } \nonumber \\
    & + \frac{ \left[ \sum_j X_j \sin \omega (t_j - \tau) \right]^2 }{ \sum_j \sin^2 \omega (t_j - \tau) } \Bigg\},
\end{align}

where $X$ is the physical variable measured at a set of times $t_j$, and $\tau$ is defined as
\begin{equation}
    \tan (2\omega \tau) = \frac{\sum_j \sin 2 \omega t_j }{\sum_j \cos 2 \omega t_j }.
\end{equation}

The success in achieving a simple statistical behaviour is not matched by a similar favourable outcome in reducing the sampling-induced noise in the results. In this sense, the Lomb-Scargle periodogram formulation often does not show large improvements compared to the classical algorithm.

In this paper, a new implementation of the periodogram, the high-pass filter (HPF) periodogram, is proposed. Aim of this new analysis method is not to replace the periodogram, as it was developed starting from complementary purposes; its main aim is to reduce as much as possible the noise introduced by the sampling, in order to be able to establish with much better precision the PSD of the analysed signal. As will be shown in the following, a better estimation of the PSD can also favour a more reliable detection of periodicities.

The paper is organized as follows: In Section \ref{Section2}, we present our new implementation of the periodogram. Section \ref{Sect:methods} details the simulations and the testing pipeline developed to evaluate the PSD estimation provided by the LS and HPF periodograms. It also explains the theoretical basis of periodicity detection using the periodogram, defines the detection threshold, and outlines the pipeline for injecting periodic signals into our simulations and using the periodograms for periodicity identification. In Section \ref{Sect:results}, we present the results from the testing pipelines. Finally, Section \ref{Conclusions} summarizes the conclusions of this paper.

\section{The high-pass filter periodogram} 
\label{Section2}

The algorithm proposed in this paper aims to reduce the sampling-induced noise in the periodogram and is the result of considerations concerning the origin of the noise. The power of the analysed signal returned by the periodogram at a given frequency $f$ is transferred to other frequencies, in a way that is determined by the irregular sampling. The higher the power in the signal, the more significant the peaks induced at other frequencies by the sampling. In the common astronomical case where the signal can be approximated, within a large range of frequencies, by red noise, i.e. PSD($f$) $\propto f^{-2}$, or at least with flicker noise, i.e. PSD($f$) $\propto f^{-1}$, the low frequency components of the signal carry much more power than the high frequency ones. Therefore, the power aliased from low frequencies is more substantial compared to that aliased from high frequencies. This explains why the low frequency values of the periodogram are marginally affected by the sampling, while the high ones can be heavily dominated by noise.

Based on these considerations, we developed an approach that applies a frequency-dependent high-pass filter to the signal before calculating the periodogram value at a given frequency\footnote{The high-pass filter periodogram tool (DOI: \href{https://doi.org/10.5281/zenodo.13917829}{10.5281/zenodo.13917829}) is available for download on GitHub: \url{https://github.com/ealruiz/HPF_Periodogram}.}. For each frequency $f$, we compute the periodogram of the signal after suppressing, in the time domain, the variability components of frequencies lower than $f$ through the data de-trending algorithm developed by \citep{2002A&A...390..407V}. The detailed steps of the process are outlined as follows:
\begin{itemize}
    \item For a given frequency $f$, the original light curve is averaged over time bins of size $\Delta$t=1/$f$.
    \item The averaged points are fitted with a spline curve, which is resampled to match the original data. This provides an estimate of the long-term trend on timescales longer than $\Delta$t.
    \item The resampled spline is subtracted from the original data, yielding a de-trended light curve with the same sampling of the original data.
    \item The periodogram at frequency $f$ is calculated on the de-trended light curve.
\end{itemize}
\noindent This procedure is repeated for all the frequencies at which the periodogram is computed. By removing the influence of lower frequencies, this approach significantly reduces the noise transferred from low to high frequencies.

\begin{figure}
    \centering
    \includegraphics[width=\linewidth]{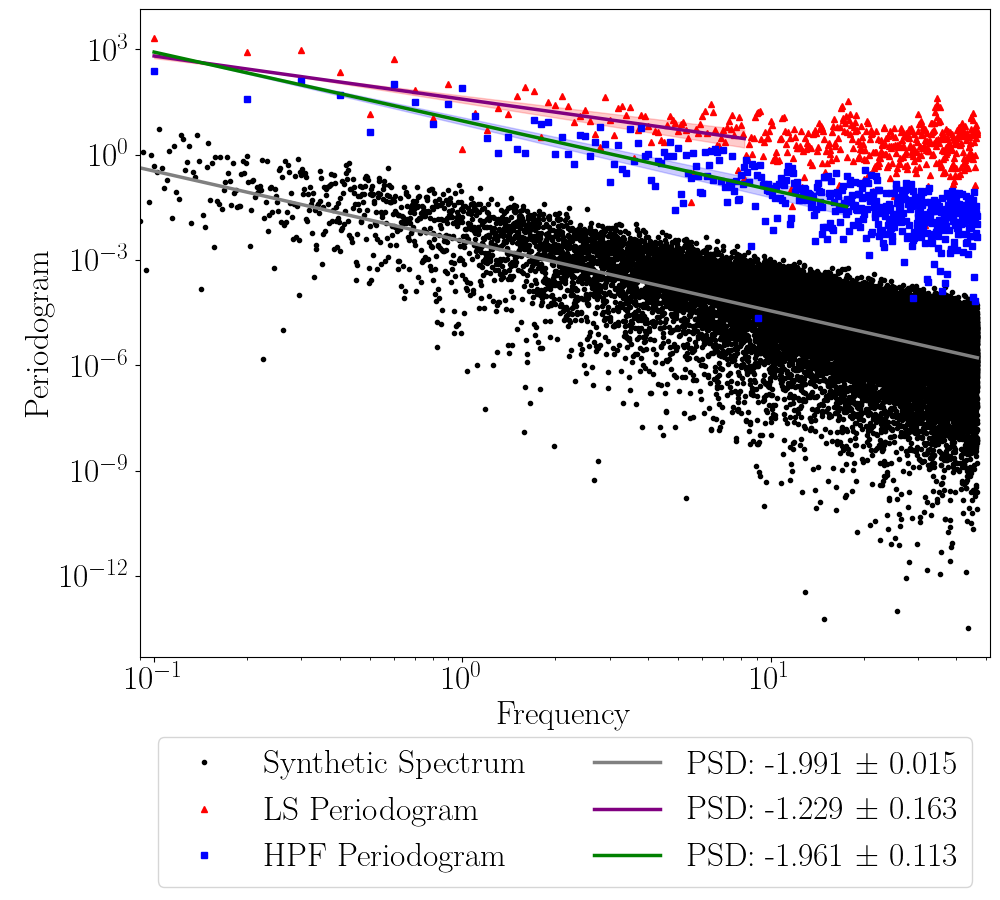}
    \caption{Synthetic power-law spectrum (black dots), along with the Lomb-Scargle (LS, red triangles) and high-pass filter (HPF, blue squares) periodograms calculated from the synthetic signal, which corresponds to the synthetic spectrum. The grey, purple, and green lines represent the linear fit to the synthetic spectrum, LS periodogram, and HPF periodogram, respectively. At high frequencies, the LS periodogram flattens due to noise introduced by the sampling, while the HPF periodogram more closely matches the synthetic power spectral density (PSD).}
    \label{fig:PSD_LSvsHPF}
\end{figure}

\begin{figure*}
    \centering
    \includegraphics[width=8.75cm]{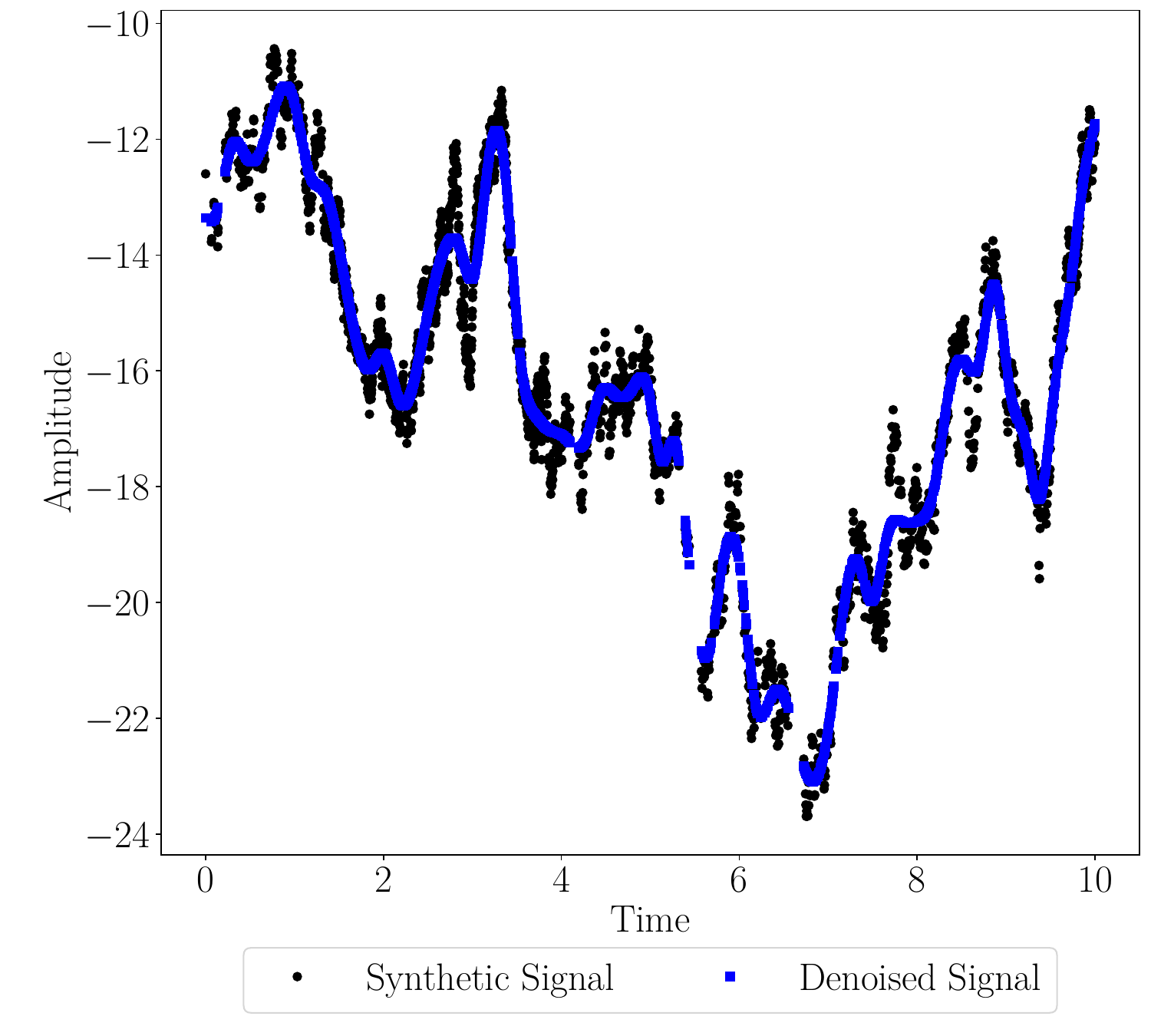}
    \includegraphics[width=8.75cm]{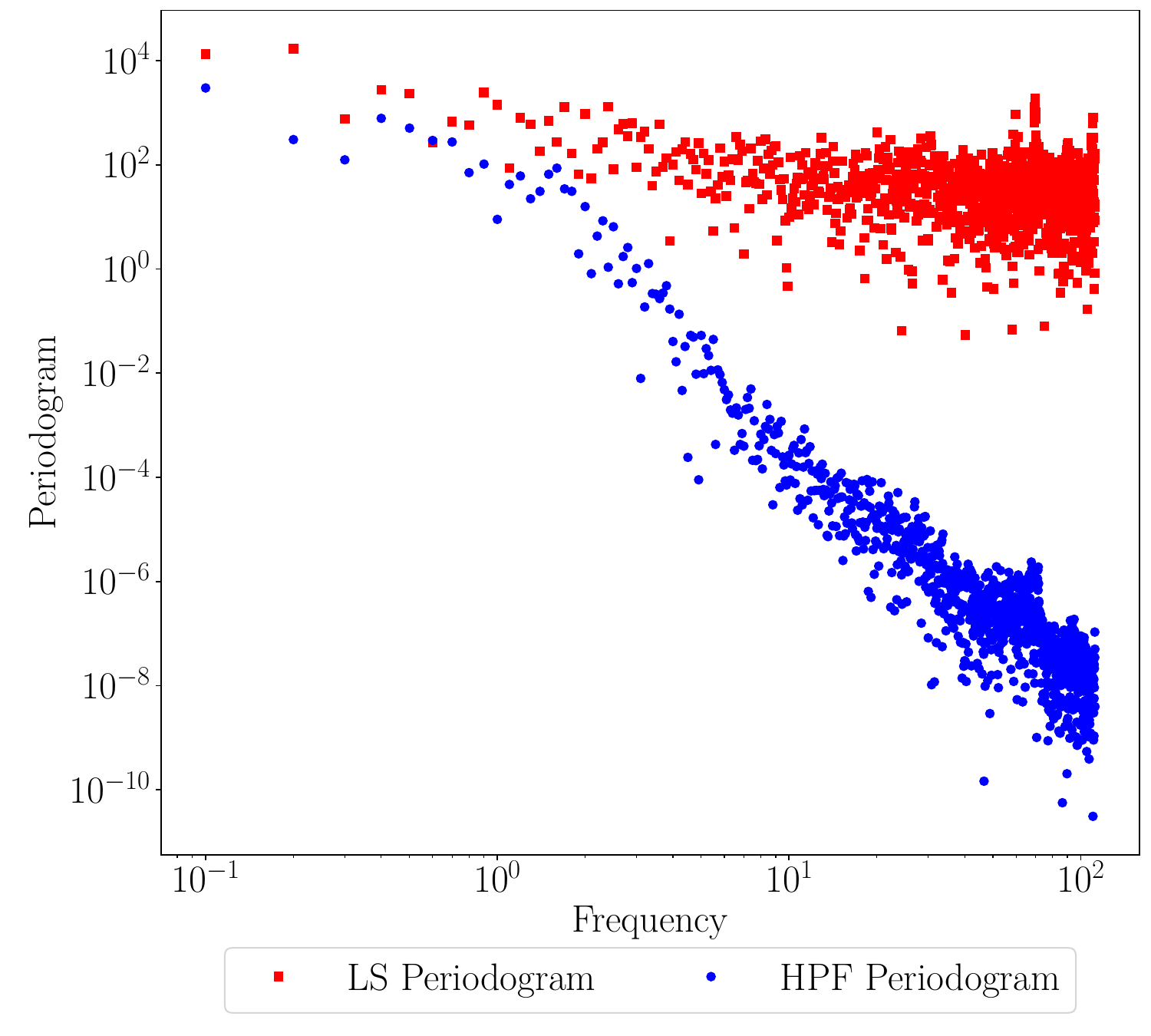}
    \caption{Left: Synthetic signal (black points) and de-noised signal (blue squares). The de-noised signal is obtained by binning the synthetic signal in intervals of 1/6 of the time unit. Right: LS and HPF periodograms calculated from the de-noised signal. At high frequencies, de-noising causes a decrease in the power of the HPF periodogram, a decrease that is not observed in the LS periodogram, where noise dominates the values at high frequencies.}
    \label{fig:LSvsHPF_denoised}
\end{figure*}

Two examples of the HPF results are shown in Fig. \ref{fig:PSD_LSvsHPF} and \ref{fig:LSvsHPF_denoised}. In the first case, a synthetic power-law spectrum was generated with slope -2, and then transformed into a synthetic light curve through the method proposed by \cite{Timmer1995}, using a moderately uneven sampling. The flattening at high frequencies of the LS periodogram applied to the light curve leads to an important underestimation of the PSD slope of the signal, while the HPF periodogram returns a pretty faithful representation of it. In Fig. \ref{fig:LSvsHPF_denoised} a synthetic light curve is generated in the same way as explained above, and then is de-noised by binning the signal into intervals of 1/6 of the time unit, and fitting a spline curve over the binned flux density values. A spline curve is fitted over the binned flux density values. The PSD of a de-noised light curve should show a dramatic drop in power at high frequency, which is not visible in the LS periodogram because of the noise, while it is very clear in the HPF version.

\section{Methodology}
\label{Sect:methods}

We have introduced above a novel procedure for estimating the PSD of a given signal. In this section, we outline the various tests conducted to assess the performance of our algorithm and compare it with results obtained using the periodogram. 

As previously mentioned, the periodogram is widely employed to characterize the time-variability of signals, as fitting the periodogram slope provides an estimate of the signal's Power Spectral Density (PSD). Therefore, our initial step involves evaluating the performance of our new periodogram implementation in retrieving the PSD of a sample of signals.

For this purpose, we have devised the following testing pipeline for a given sampling:
\begin{itemize}
    \item First, generate a sample of simulated signals following a specific power law spectrum $S(\omega)\sim \omega^{-\beta}$, where $\omega=2\pi f$ and $\beta$ represents the PSD slope of the signal. We utilize the algorithm described by \cite{Timmer1995} for generating each simulated signal.
    \item Add random normal noise centered at 0 with a standard deviation (std) equal to a fraction of the signal's standard deviation, to each simulated signal. We set this fraction to $3\%$, which seems a reasonable level for generic astronomical datasets. For a discussion of the effect of different noise levels on the periodogram estimates, refer to Appendix \ref{Ap:NoiseEffect}.
    \item Compute the periodogram for each simulated signal using both the Lomb-Scargle approach and the new implementation presented in this work.
    \item Retrieve the PSD of each signal by fitting the slope of the periodogram using the model $\log P(\omega) = -b\log \omega + a$.
\end{itemize}

The slope obtained for each simulated signal should ideally match the specified PSD slope $\beta$. However, since white noise has been added to the data to make them more realistic, the periodogram flattens at high frequencies, when the variability due to the noise becomes comparable or higher than the one due to the simulated signal. These frequencies need to be discarded before fitting the signal's PSD slope. To handle this automatically across our extensive sample of simulated signals, manual noise level inspection for each simulation is impractical. Therefore, we implemented an algorithm to detect the change point in the periodogram (i.e., the frequency where noise dominates over the signal contribution to the periodogram) for each simulated signal, following the bottom-up method (see for instance \citealt{Keogh2001} or \citealt{Fryzlewicz2007}) implemented in the \texttt{ruptures} Python library (see \citealt{Truong2020})\footnote{A discussion on the effect of using different segmentation methods on the results of our analysis in presented in Appendix \ref{Ap:NoiseFreq}.}. By using this algorithm, we identify the frequencies at which the noise dominates over the signal in the periodogram, which are then discarded in the computation of the periodogram slope (that is, the estimated PSD of the signal).

\begin{figure}[h!]
\centering
   \includegraphics[width=9cm]{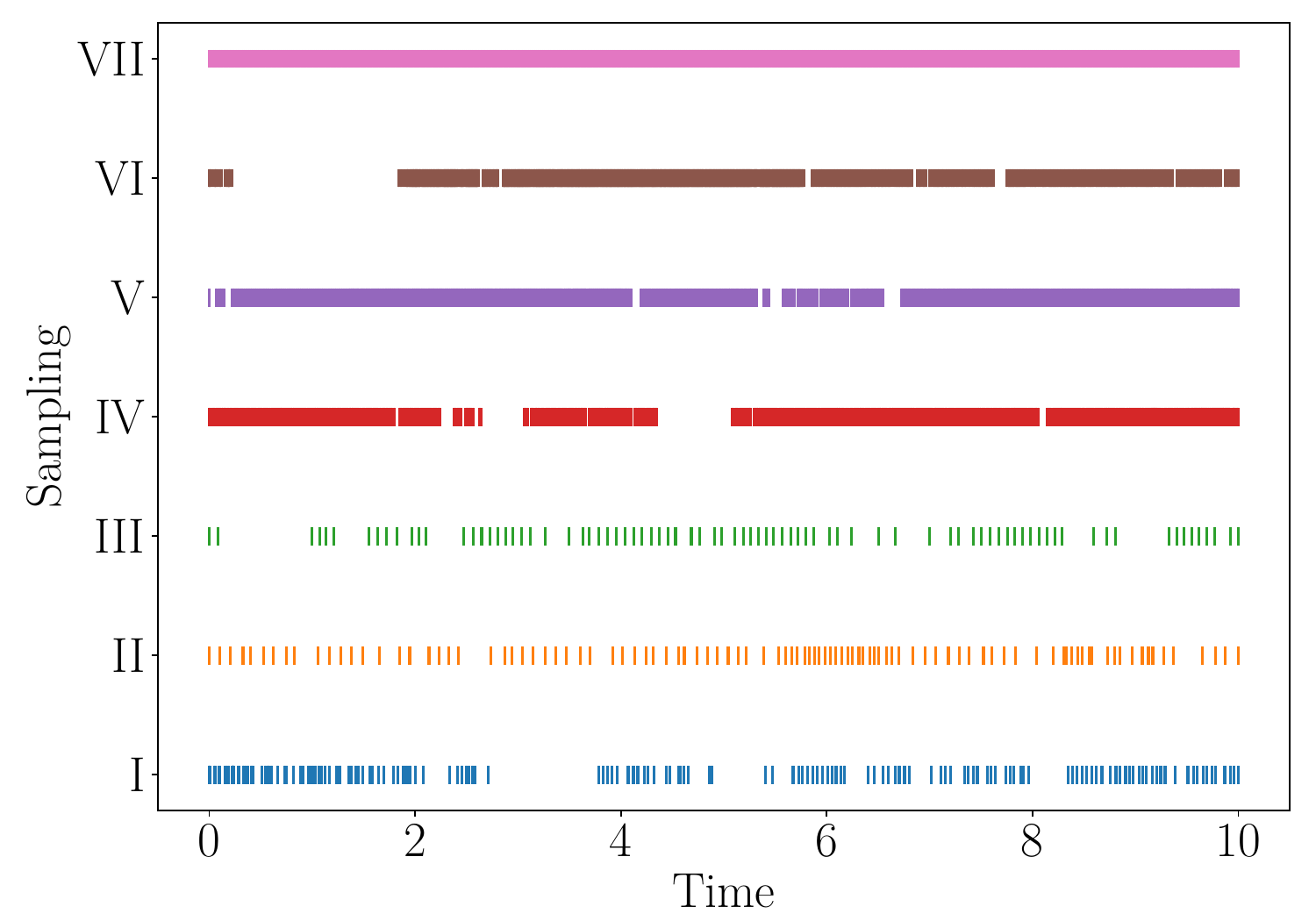}
   \caption{Visualization of the seven selected samplings.}
   \label{fig:samplings}
\end{figure}

Seven samplings were selected from light curves of different projects (the IRA Blazar Monitoring program, see \citealt{2007A&A...464..175B}; the F-GAMMA program, see \citealt{2016A&A...596A..45F, 2019A&A...626A..60A}; the US Navy’s extragalactic source monitoring program, see \citealt{1987ApJS...65..319F, 2001ApJS..136..265L}) for the generation of the synthetic light curves used in the pipeline. The seven samplings, briefly described below, and shown in Fig. \ref{fig:samplings}, were chosen because of the differences in their main characteristics, which can be summarised in the number of data-points, the presence of large gaps, the regularity of the cadence excluding the gaps:
\begin{enumerate}[label=\Roman*.]
    \item relatively low number of data-points (i.e. 246 points), highly irregular sampling, some large gaps in the data; 
    \item low number of data-points (100), mildy irregular sampling, no large gaps; 
    \item low number of data-points (83), irregular sampling, a few large gaps in the data; 
    \item relatively large number of data-points (935), sampling mostly mildy irregular, some large gaps in the data; 
    \item large number of data-points (2240), sampling mostly mildy irregular, some small gaps in the data; 
    \item large number of data-points (2774), sampling mostly mildy irregular, a single very large gap in the data; 
    \item intermediate number of data-points (500), even sampling. 
\end{enumerate}

To ensure a diverse sample of signals, we generate 333 signals for 11 PSD slope values, $\beta=\{2.5, 2.4, 2.3, 2.2, 2.1, 2.0, 1.9, 1.8, 1.7, 1.6, 1.5\}$. This results in a total of 3663 simulated signals spanning a wide range of PSDs for each sampling.

\subsection{Periodicity Detection}

The periodogram is also valuable for detecting periodicities in signals. Therefore, we can further evaluate the performance of the new periodogram implementation by assessing its capability to detect periodicities in a sample of simulated periodic signals.

To determine a periodicity, we employ the statistical analysis method developed in \citet{ScargleII1982}. Let us start from assuming that the analysed signal $X$ is pure noise; in this case, the power at a given frequency is exponentially distributed, and the probability distribution for $Z=P_X(\omega)$ is 
\begin{equation}
p(z) dz= \mathrm{Pr}(z<Z<z+dz)=\exp(-z) dz.
\end{equation}
The statistical significance of a periodogram value z estimated at a single frequency is Pr$(Z>z) = \exp(-z)$.

If we now explore a set of $N$ frequencies, assuming that the periodogram values $\{P(\omega_n), n=1,2,\ldots,N\}$ are independent random variables, we can estimate the probability to find a value equal or higher than the maximum one in the periodogram, $z_{max}$:
\begin{equation}
    Pr\{z_{max}>z\}=1-[1-\exp(-z)]^N.
\end{equation}

We can now exploit this information to set a detection threshold $z_0$, i.e. a limit in the power of the periodogram calculated at a given frequency above which we can claim the detection of a periodicity in the signal:
\begin{equation}
    z_0 = -\ln \left[1-(1-p_0)^{1/N} \right],
    \label{eq:detect_threshold}
\end{equation}
where $N$ is the number of frequencies inspected to identify the maximum power of the periodogram, and $p_0$ denotes the false alarm probability, that is, the  likelihood of incorrectly claiming detection of periodicity when the power $P(\omega_n)$ of the periodogram at a certain frequency exceeds $z_0$. We set $p_0$ to 0.01.

However, the periodogram of a generic signal does not adhere to the probability distribution of pure noise signals with zero mean and constant variance, which is the case explored in \citet{ScargleII1982}. Assuming that the PSD of the signal can be modelled as a power-law, to make our statistical analysis applicable we have to compute a normalised periodogram, denoted as $P_N$, by dividing the periodogram by the power law fit obtained from the slope estimation, that is, $P_N(\omega) = P_X(\omega) / 10^{(-b\log \omega + a)}$, where $P_X(\omega)$ is the original periodogram, and $(-b\log \omega + a)$ corresponds to the model fitted to the periodogram. This normalization is performed after discarding frequencies where noise dominates over the signal contribution to the periodogram. Finally, the normalised periodogram is scaled to an average value of 1, to keep the equations above valid.

Taking all this into consideration, we implemented an alternative testing pipeline designed to identify periodicities. For each of the samplings, the steps are as follows:
\begin{itemize}
    \item Generate a sample of simulated signals following the first two steps of the testing pipeline.
    \item Add a periodic signal to each simulated signal, computed as \( \{Y_j = Y_0 \sin(\omega_0 \cdot t_j), j=1,2,\ldots,N\} \), where \( t_j \) represents the set of times of the original signal. \( Y_0 \) denotes the amplitude of the periodic signal, calculated as a fraction of the standard deviation of the original signal. \( \omega_0 = 2\pi / T_0 \) is the frequency of the periodic signal, determined by choosing a period \( T_0 \) that is a fraction of the duration of the signal.
    \item Compute the periodogram for each simulated periodic signal using both the classical approach and the new implementation presented in this work.
    \item Utilize the previously implemented automatic detection method to identify and discard frequencies where noise dominates over the signal contribution to the periodogram. Then, retrieve the PSD of each periodic signal by fitting the slope of the periodogram.
    \item Calculate the normalised periodogram \( P_N \) for both the classical and the new implementation, considering frequencies within the range not dominated by the noise contribution to the periodogram.
    \item Evaluate the power of the normalised periodogram at the \( N \) filtered frequencies and claim detection of periodicity if the power at frequency \( f \) exceeds the detection threshold \( z_0 \) given by equation \ref{eq:detect_threshold}.
    \item Calculate the detection efficiency of both periodogram implementations, defined as the fraction of simulations for which a periodicity is detected at a frequency $f$ matching the one of the injected periodic signal \( f_0 = 1/T_0 \).
\end{itemize}

\begin{figure}
    \centering
    \includegraphics[width=\linewidth]{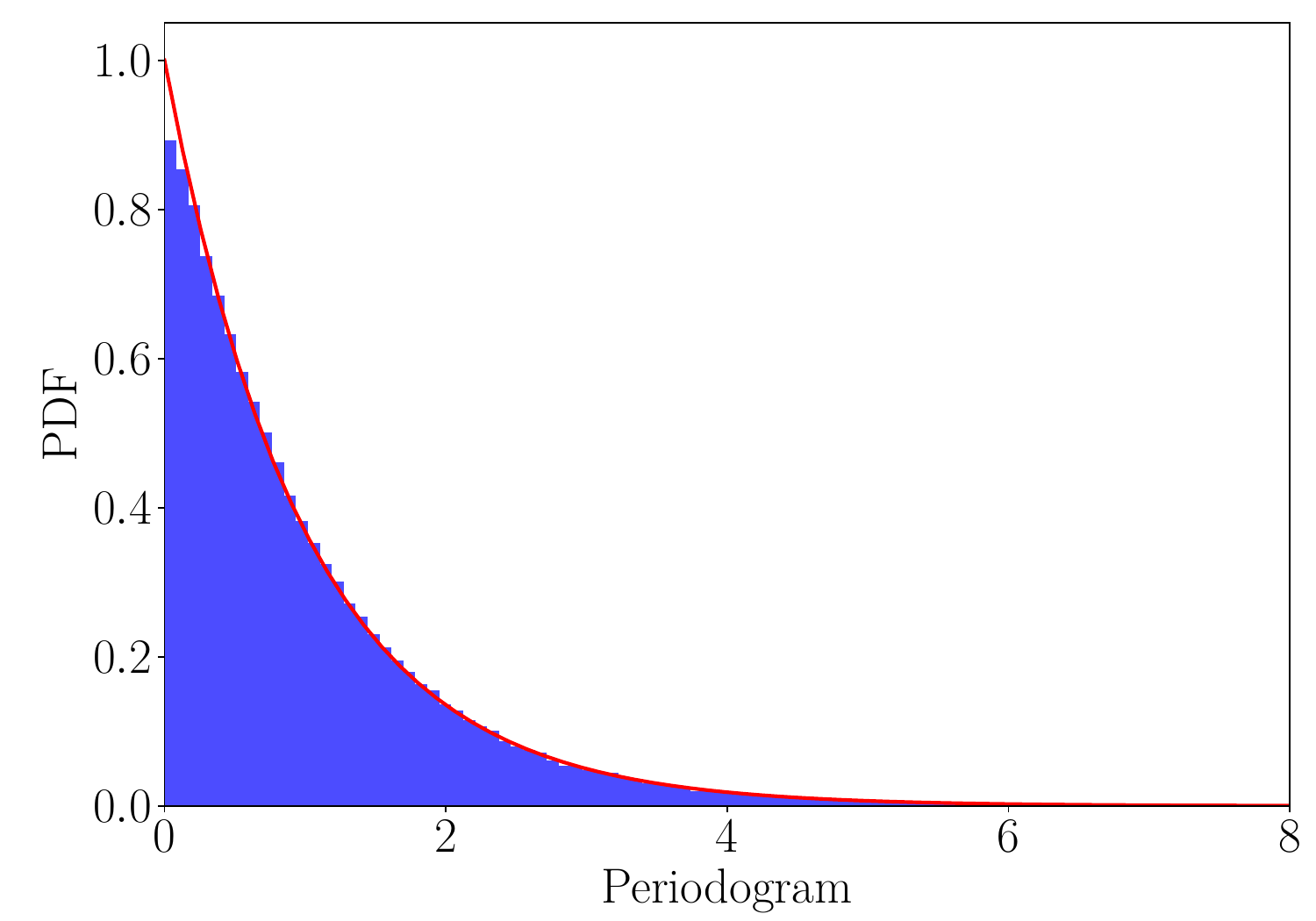}
    \caption{Histogram of the normalized periodogram values obtained from a sample of 1111 simulated signals with sampling V and $\beta = 2.0$. The red line represents the theoretical probability distribution of the normalized periodogram, $\exp(-z)$.}
    \label{fig:PDF_Periodogram}
\end{figure}

The hypothesis that the periodogram values follow an exponential distribution is tested using the $\chi^2$ and Kolmogorov-Smirnov (KS) tests. To evaluate this hypothesis, we simulate 1111 signals with a PSD($F$)$\propto f^{-2}$. We then calculate the normalized high-pass filter (HPF) periodogram, scaled to an average value of 1. The histogram of the normalized periodogram values should conform to the theoretical $\exp(-z)$ probability distribution, as illustrated in Figure \ref{fig:PDF_Periodogram}. The $\chi^2$ test applied to the periodogram values yields a low $\chi^2$ statistic of 0.019 with a P-value of 0.999, while the KS test gives a statistic of 0.053 with a P-value of 0.998. These results provide strong evidence that the periodogram values, P($\omega$), follow an $\exp(-z)$ distribution.

We follow this pipeline for seven selected samplings. To create a comprehensive set of periodic signals, we first generate 1111 signals for three PSD values, $\beta=\{2.5,2.0,1.5\}$, and then add periodic signals of amplitudes $X_0$ of $0.3$, $1$ and $3$ times the std of the original signal, and period of $\{0.01,0.05,0.1,0.25,0.4\}$ times the duration of the signals. Consequently, for each sampling method, we generate 3333 simulated signals, with each signal independently injected with 15 periodic signals of varying amplitudes and periods.

\section{Results and discussion}
\label{Sect:results}

In this section, we present the results obtained from the implemented testing pipeline. Due to the extensive number of simulated light curves, we conduct a statistical analysis on the combined samples across different simulated PSDs for each sampling method. This approach allows us to compare the performance of both the classic and the new periodogram implementations across various signal characteristics.

\begin{table}[h!]
    \centering
    \caption{Median and uncertainties of the discrepancy between the true PSD $\beta$ of a sample of simulated signals and the PSD estimated from the classic (LS) and the high-pass filer (HPF) periodograms for all testing samplings. The lower and upper uncertainties correspond to the $1\sigma$ dispersion of the PSD discrepancy distribution below and above the median value, respectively.}
    \begin{tabular}{lcc}
        \toprule
        Sampling & \multicolumn{2}{c}{\((\beta - \text{b})/\beta\)} \\
         & LS & HPF \\
        \midrule
        I & \(0.85^{+0.04}_{-0.31}\) & \(0.29^{+0.25}_{-0.29}\) \\
        \addlinespace[0.5em]
        II & \(0.46^{+0.14}_{-0.26}\) & \(0.02^{+0.22}_{-0.20}\) \\
        \addlinespace[0.5em]
        III & \(0.87^{+0.10}_{-0.28}\) & \(0.05^{+0.25}_{-0.23}\) \\
        \addlinespace[0.5em]
        IV & \(0.42^{+0.15}_{-0.27}\) & \(0.07^{+0.10}_{-0.11}\) \\
        \addlinespace[0.5em]
        V & \(0.44^{+0.15}_{-0.24}\) & \(0.06^{+0.06}_{-0.07}\) \\
        \addlinespace[0.5em]
        VI & \(0.47^{+0.11}_{-0.18}\) & \(0.05^{+0.06}_{-0.07}\) \\
        \addlinespace[0.5em]
        VII & \(0.11^{+0.22}_{-0.21}\) & \(0.06^{+0.11}_{-0.12}\) \\
        \bottomrule
    \end{tabular}
    \label{tab:beta_comparison}
\end{table}

\begin{figure*}[htp]
\centering
   \includegraphics[width=17.5cm]{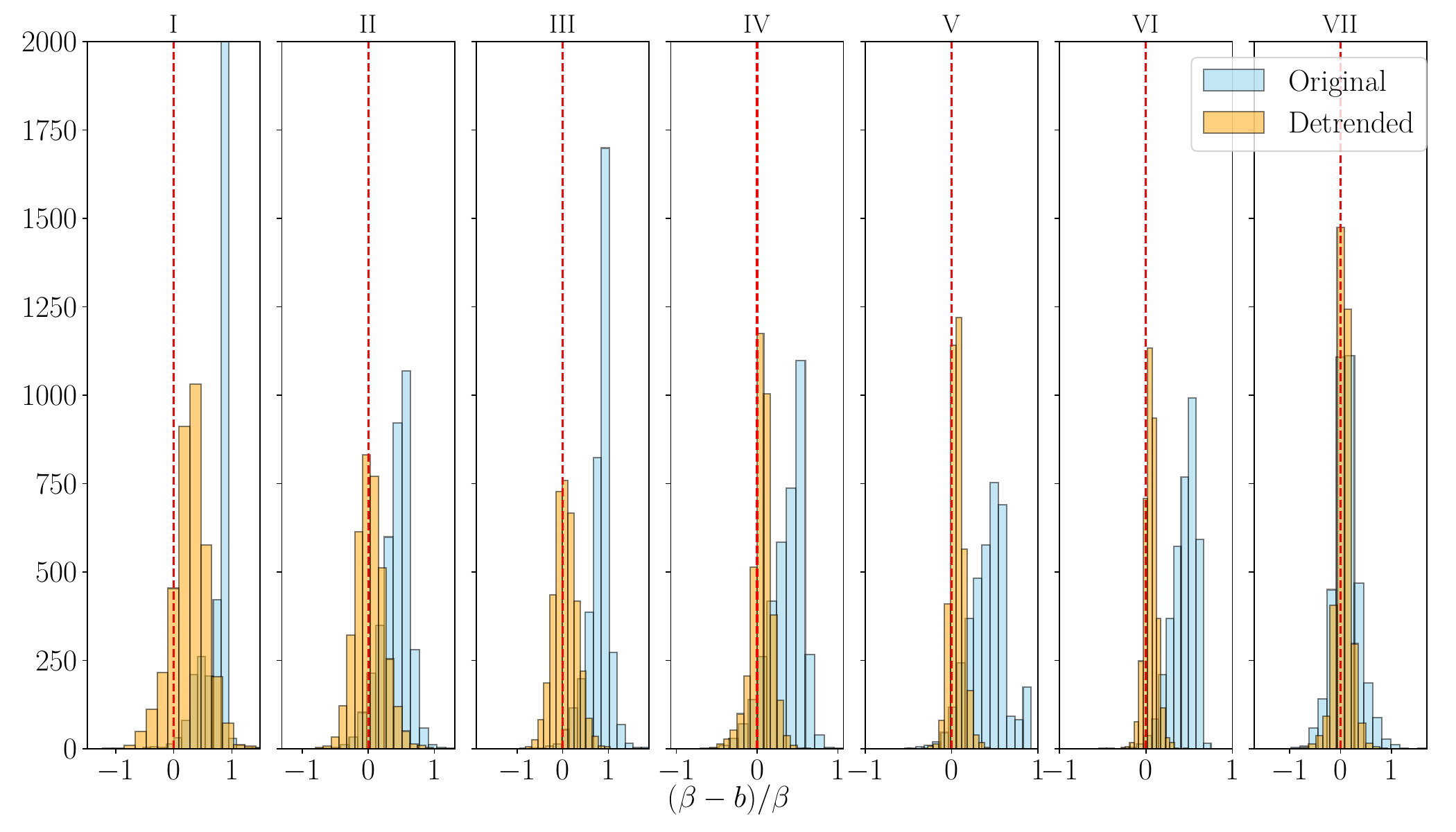}
   \caption{Statistics on PSD estimates retrieved from the periodogram of a sample of simulated signals, comparing the classic (in blue) and new (in orange) implementations, across seven different testing samplings.}
   \label{fig:PSD_Fitting}
\end{figure*}

In Table \ref{tab:beta_comparison}, we provide statistics on the discrepancies between the estimated PSD values obtained using both the classic and new implementations of the periodogram, compared to the imposed $\beta$ values during signal simulation across seven different samplings. Figure \ref{fig:PSD_Fitting} displays histograms of these discrepancies, visually illustrating the performance comparison between the periodogram implementations.

We observe a significant improvement in PSD estimation using our new implementation of the periodogram compared to the classic approach across all samplings, except for the even sampling where both versions provide accurate PSD estimates as expected. Numerically, the discrepancies obtained with the new periodogram range around $5-10\%$ for most samplings, with the worst explored sampling showing discrepancies around $30\%$. In contrast, PSD estimates retrieved with the classic periodogram exhibit discrepancies ranging from $40-90\%$.

The $5-10\%$ discrepancy from the PSD of the simulated signal estimated with the new periodogram implementation can be primarily attributed to two sources of uncertainty\footnote{A discussion on the effect of the amplitude of the noise injected in the simulations, and the use of different segmentation methods in the PSD estimates derived using the periodogram, is presented in Appendices \ref{Ap:NoiseEffect} and \ref{Ap:NoiseFreq}, respectively.}:
\begin{itemize}
    \item Noise in the signal: the amplitude of the white noise added to the simulated signal has obviously an impact on the capability of the algorithm to retrieve the signal's characteristics.
    \item Automated identification of noise-dominated frequencies: We automatically identify the frequencies in the signal's frequency domain where the noise contribution dominates over the signal in the periodogram, using a Bottom-Up segmentation method. This automation is necessary for robust statistical analysis to evaluate the performance of both periodogram implementations. However, when analyzing signals from astronomical observations, manual identification of these frequencies through rigorous analysis is recommended to improve the accuracy of the estimates.
\end{itemize}

\subsection{Periodicity Detection}
In Section \ref{Sect:methods}, we present a pipeline for identifying periodicity within a signal by examining the value of the normalised periodogram at specific frequencies in the signal's frequency domain. Specifically, a periodicity is claimed at a period $T=1/f$ if the normalised periodogram measured at the frequency $f$ in the signal's frequency domain exceeds the threshold level given by Eq. \ref{eq:detect_threshold}, with a false alarm probability $p_0=0.01$.

\begin{figure*}[htp]
\centering
   \includegraphics[width=18cm]{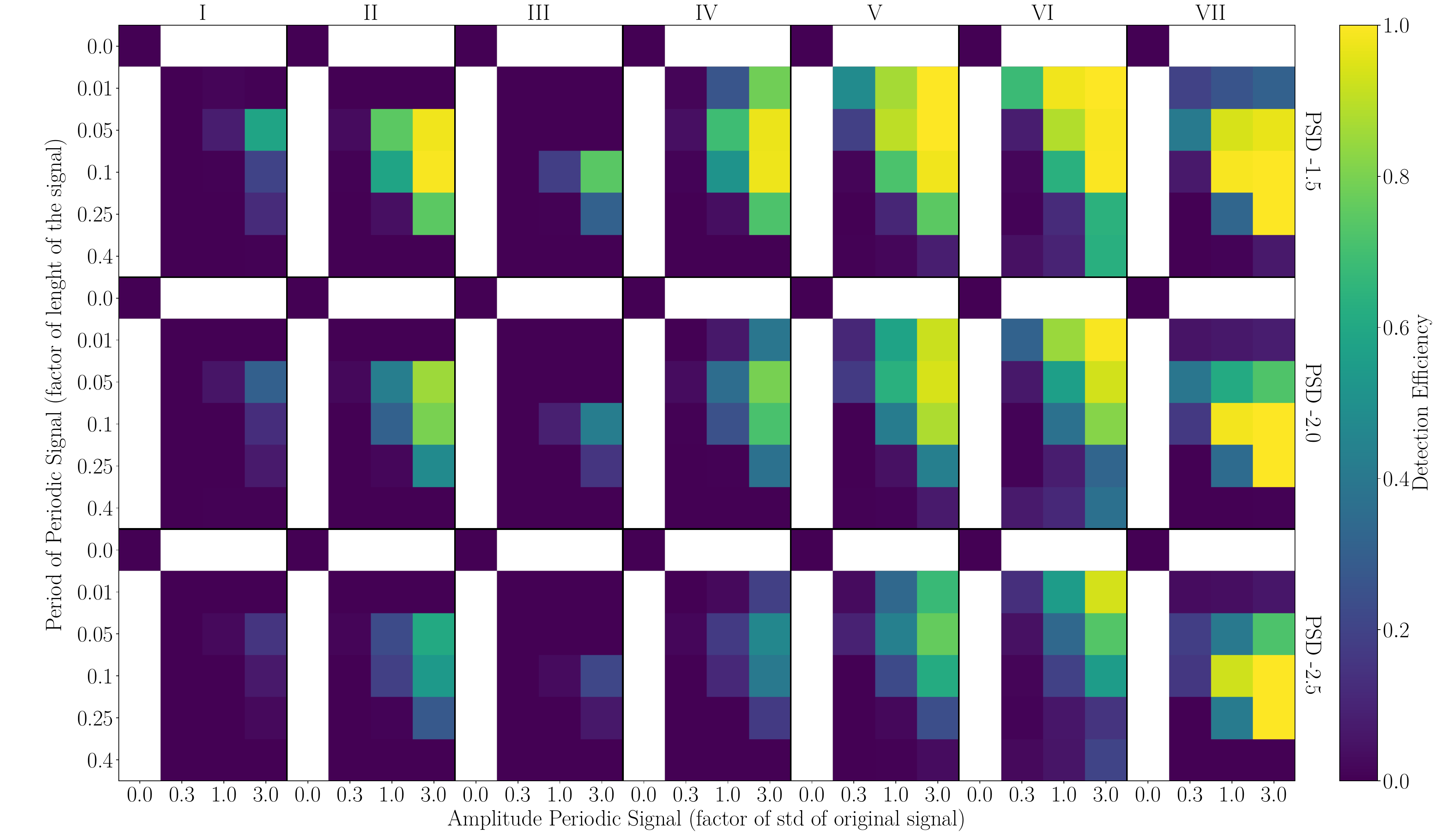}
   \includegraphics[width=18cm]{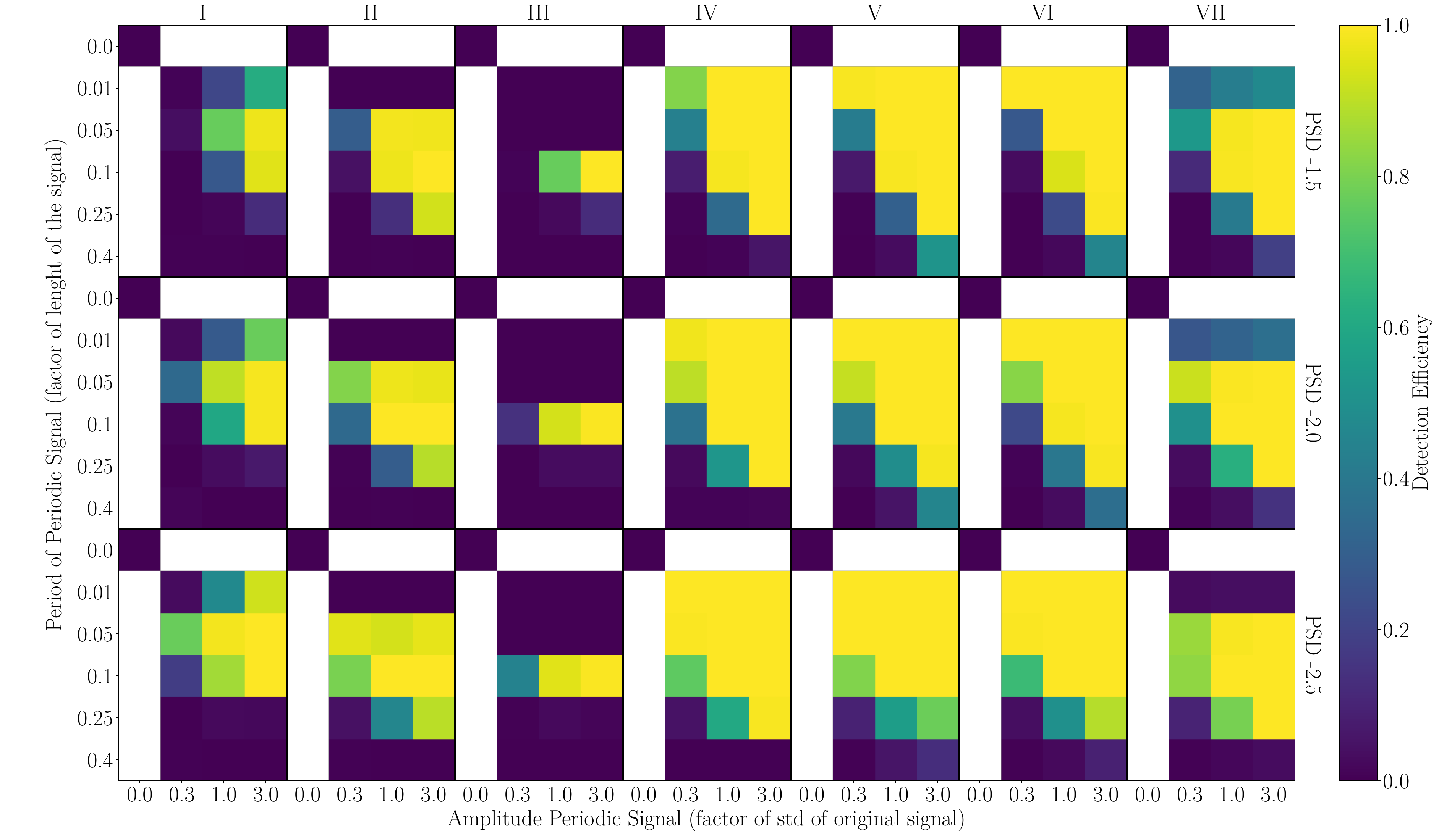}
   \caption{Detection Efficiency (DE) of the Lomb-Scargle (top) and HPF (bottom) periodogram implementations, shown in the colour map. The DE is derived from a sample of simulated signals, further modified by adding a periodic signal with varying amplitudes (x-axis values) and periods (y-axis values), for each combination of sampling (columns) and signal PSD (rows). The DEs for the sample of simulated signals without a periodic signal corresponds to the 0 amplitude and 0 period values for each sampling and PSD.}
   \label{fig:DE_classic_and_new}
\end{figure*}

We follow this criterion by exploring the normalised periodogram of a comprehensive set of simulated signals, each injected with periodic signals of varying amplitudes and periods. This allows us to evaluate the Detection Efficiency (DE) of both the LS periodogram and the new version proposed in this work, as shown in Figure \ref{fig:DE_classic_and_new}. The DE is defined as the ratio between the number of simulations in which we detect periodicity at the expected frequency (i.e., that of the injected periodic signal) to the total number of simulations. The color map in the figure illustrates the DE derived from the analysis of a sample of 1111 simulated signals, each generated for a combination of sampling and PSD, and injected with periodic signals of different amplitudes and periods, following the procedure outlined in Sect. \ref{Sect:methods}.

We observe that the DE is generally higher with the high-pass filter implementation of the periodogram compared to the Lomb-Scargle version, while the DE for even sampling is similar for both implementations, as expected. The improvement in the HPF periodogram implementation can be attributed not only to the motivations behind its development, but also to the enhanced estimation of the signal's PSD provided by this implementation\footnote{Refer to Appendix \ref{Ap:PeriodicitySlope} for a discussion on the effect of the injected periodic signal on the PSD estimates provided by the periodogram.}. This improved PSD estimation affects the values of the normalised periodogram, thus contributing to the higher DE. Additionally, we observe more consistent Detection Efficiency values with the High-Pass Filter implementation across different PSDs of the simulated signals, whereas the DE from the Lomb-Scargle periodogram generally decreases as the PSD becomes more negative.

When examining different samplings, we observe challenges in detecting periodicity at very low frequencies, especially at periods of 0.4 times the duration of the signal. Specifically, for the most complex sampling patterns, I and III, detecting periodicity becomes more challenging for periods longer than one-tenth of the observation duration. This difficulty arises because the limited number of cycles in the periodic signal weakens the imprint of the periodicity on the periodogram. Detection also becomes more difficult as the amplitude of the periodic signal decreases, due to a weaker contribution to the periodogram values.

A decrease in DE is also noticeable at high frequencies (periods of 0.01, and occasionally 0.05, times the duration of the signal) for poorer samplings. Upon closer inspection of the normalised periodograms, the reduced DE in these cases appears linked to noise-dominated frequencies, at which the periodogram flattens due to the white noise added to the data, being filtered out.

To estimate the PSD and compute the normalised periodogram, we automatically identify the frequencies in the signal's frequency domain where noise dominates over the signal contribution to the periodogram. If the frequency of the sought periodicity falls within this filtered frequency range, the periodicity will not be detected, thereby lowering the DE. This explains the increased DE for better samplings, as the noise primarily affects higher frequencies, allowing us to retain the frequencies where we expect periodicity.

Finally, to assess the robustness of periodicity detection, we explore the total number of periodicities claimed and evaluate the number of False Detections, on average, for each combination of sampling, PSD, and amplitude and period of the injected periodic signal. A False Detection is defined as a periodicity claimed at a frequency different from that of the injected periodic signal.

\begin{figure*}[htp]
\centering
   \includegraphics[width=18cm]{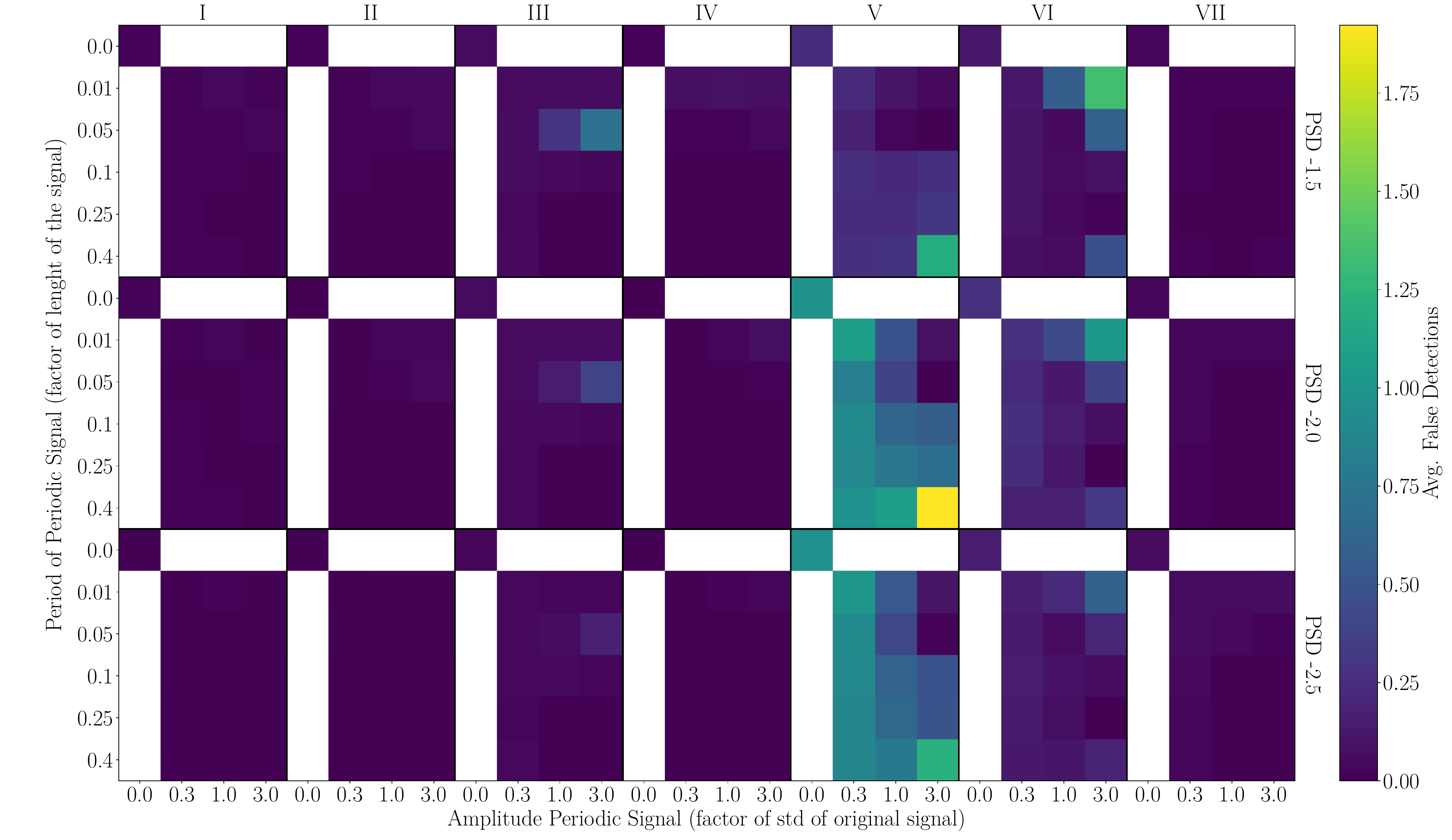}
   \includegraphics[width=18cm]{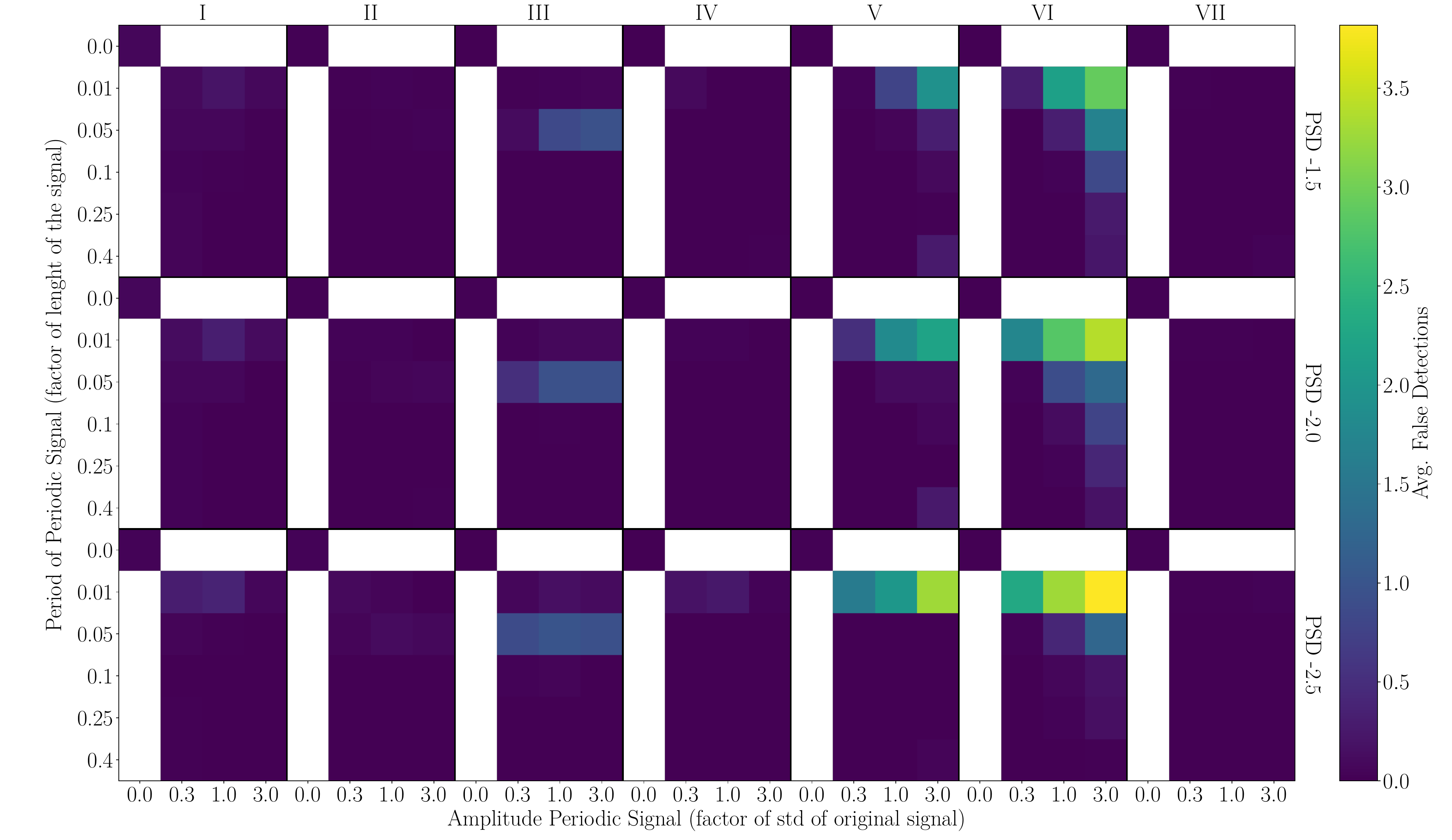}
    \caption{False Detections obtained from the Lomb-Scargle (top) and HPF (bottom) periodogram implementations, shown in the colour map. The data is derived from samples of simulated signals generated for each combination of sampling (columns) and signal PSD (rows), each modified by adding a periodic signal with varying amplitudes (x-axis values) and periods (y-axis values). The values for the sample of simulated signals without a periodic signal correspond to the 0 amplitude and 0 period values for each sampling and PSD.}
    \label{fig:FD_classic_and_new}
\end{figure*}

Figure \ref{fig:FD_classic_and_new} presents the average number of False Detections for each sample of simulations. In the case of even sampling, both the HPF and LS periodogram implementations successfully avoid false periodicity detections. Additionally, at lower frequencies and amplitudes of the injected periodic signals, both the classic and new periodogram implementations generally avoid false periodicity claims.

For samplings V and VI, which involve a large number of data points but are irregularly sampled with gaps, we observe that the LS implementation detects, on average, approximately one false periodicity at lower frequencies, increasing to around two at higher amplitudes. This means that up to two additional periodicities beyond the expected one are detected by the LS method when analyzing the normalized periodogram. Notably, this issue is also present in the periodogram where no periodic signal was injected, indicating that the sampling significantly affects the periodogram values, leading to false periodicity detections. In contrast, the HPF periodogram effectively avoids false detections for these samplings, as it is noticeably less impacted by the noise introduced by the irregular sampling.

Inspecting higher frequencies (periods of 0.01 and 0.05 times the duration of the signal), we observe that both the LS and HPF periodogram implementations exhibit more false periodicity detections as the amplitude of the periodic signal increases. This occurs because the injected signal is reverberated to higher frequencies, as a high-order harmonic or to other frequencies because of aliasing. Indeed, upon closer inspection of the normalized periodograms, we notice that these false detections have periodogram values lower than the one estimated at the frequency of the injected periodicity. If periodicity claims were limited to the frequency at which the normalized periodogram is at its maximum, these detections would be disregarded.

In one particular scenario, sampling III with a period of 0.05 times the signal duration, both the LS and HPF periodograms show a 50\% to 100\% probability of a false detection, even though no periodicity is observed at the expected frequency. This stems from the limitations of this sampling (very few data points, irregular sampling with large gaps), making more challenging to estimate the signal's PSD, thereby reducing the robustness of the normalised periodogram for detecting periodicity. However, the high probability of a false periodicity detection under these extreme conditions should not be overemphasized when evaluating the performance of the periodograms, as even in this worst-case scenario, periodograms of light curves with no injected periodicity consistently did not produce false detections. Moreover, these issues can be mitigated when analyzing individual signals by manually inspecting the periodogram to better estimate the signal's PSD.

\section{Conclusions}
\label{Conclusions}

This paper introduces a novel periodogram implementation designed to reduce the noise introduced by irregular sampling in time series analysis, in order to enhance the accuracy of power spectral density estimation and periodicity detection in unevenly sampled data.

By evaluating the characterization of time-variability in a sample of simulated signals using the periodogram, we find that the high-pass filter implementation consistently reduces discrepancies between true and estimated PSD values for irregular samplings, achieving discrepancies of $5-10\%$. This underscores the method's ability to deliver more accurate PSD estimates, even under complex sampling conditions where noise is transferred from low to high frequencies.

Additionally, this new implementation enhances the detection efficiency of periodic signals across a wide range of frequencies and amplitudes. However, efficiency decreases at lower frequencies due to the limited number of cycles in the periodic signal, which weakens its imprint on the periodogram. Similarly, efficiency is reduced at lower amplitudes because of the weaker contribution of the periodic signal. On the other hand, the incidence of false periodicity detections is reduced, particularly at higher frequencies and amplitudes, further highlighting the method's reliability in distinguishing genuine periodic signals from noise. The improvement in the robustness of this results stems from this implementation's ability to maintain the expected distribution of the normalised periodogram values for pure noise. 

In conclusion, the high-pass filter periodogram is a valuable tool not only in astronomy but also in various fields that require the analysis of unevenly sampled time series data. Its ability to provide accurate PSD estimates and reliable periodicity detection enhances its potential for time series analysis. By addressing key limitations of the classic Lomb-Scargle method, this new implementation offers a complementary approach for more precise and robust time-variability analysis.

\begin{acknowledgments}
This work has been partially supported by the Generalitat Valenciana GenT Project CIDEGENT/2018/021 and by the MICINN Research Project PID2019-108995GB-C22.
    
This work has been supported by the grant PRE2020-092200 funded by MCIN/AEI/ 10.13039/501100011033 and by ESF invest in your future.
\end{acknowledgments}

\newpage

\appendix

\section{Effect of Noise on Spectral Analysis.}
\label{Ap:NoiseEffect}

The simulated signals generated from the testing pipeline proposed in Sect. \ref{Sect:methods} were injected with random normal noise centered at $0$ with a standard deviation of $3\%$ of the signal's std. This noise affects the periodogram evaluated at different frequencies in the frequency domain, contributing to the discrepancy between the estimated PSD and the true PSD of the simulated signal, as shown in Figure \ref{fig:PSD_Fitting}.

\begin{figure*}[htp]
\centering
    \includegraphics[width=16.5cm]{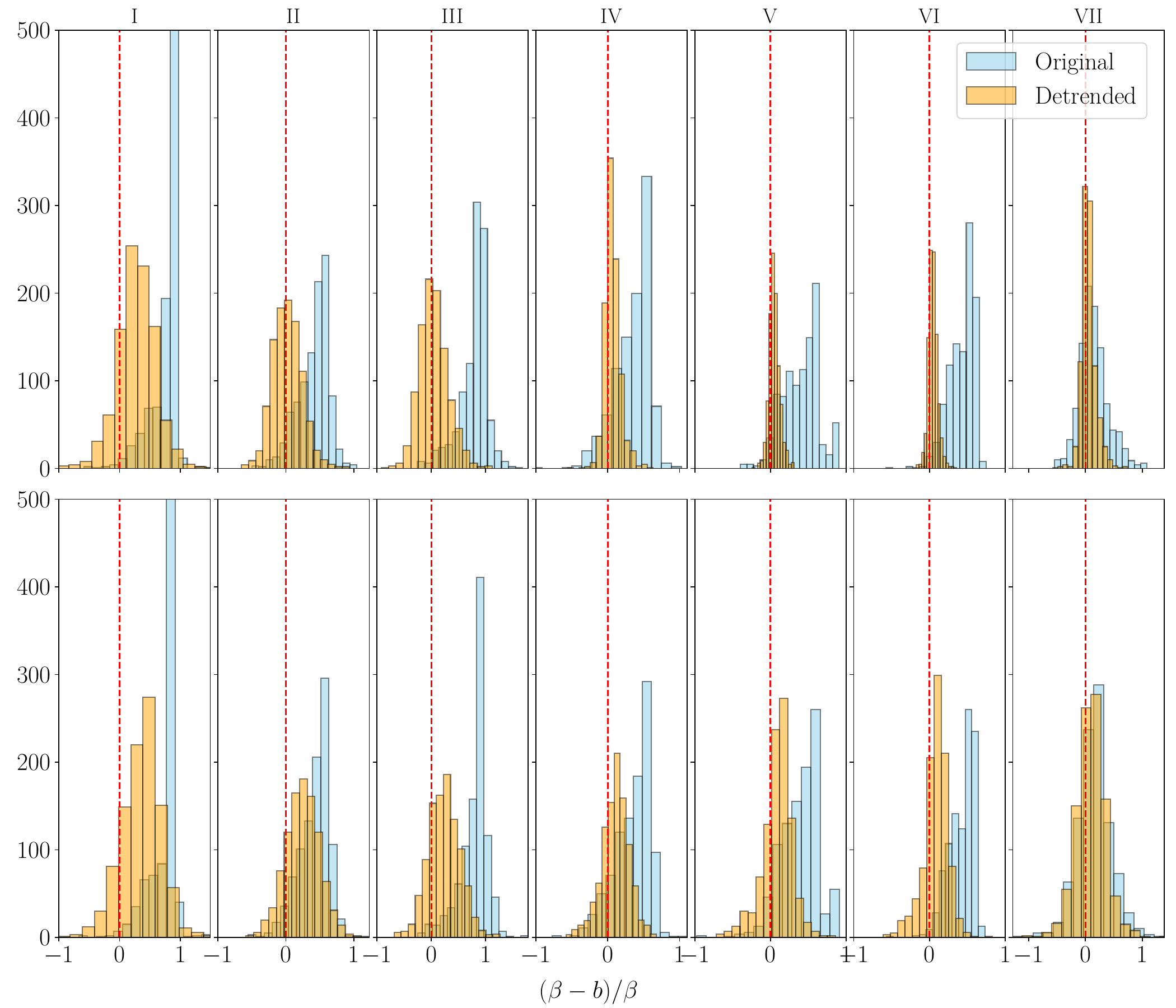}
     \caption{Statistics of the PSD estimates obtained from the periodogram of a sample of simulated signals, without injected noise (top figure) and with $0.25 \sigma$ noise (bottom figure), using the classic (blue) and new (orange) implementations for each sampling tested.}
     \label{fig:PSD_diffnoise}
\end{figure*}

In Figure \ref{fig:PSD_diffnoise}, we present the results from a similar analysis, for a reduced sample of simulated signals (we generate 333 signals with three PSD values, $\beta=\{2.5,2.0,1.5\}$), both without noise and with noise centered at $0$ with a standard deviation of $25\%$ of the signal's std. We observe a clear effect of the increased noise on the statistics of the PSD estimates, slightly degrading the performance of both the classic and new implementations of the periodogram across the seven sampling methods studied. However, our proposed periodogram implementation consistently provides better estimates of the signals' PSD, even after the injection of substantial noise.

\section{Effect of periodicity on Spectral Analysis.}
\label{Ap:PeriodicitySlope}

Section \ref{Sect:methods} provides pipelines for generating samples of simulated signals and retrieving PSD estimates,as well as detecting signal periodicities. Section \ref{Sect:results} presents the results, comparing the performance of both the classic and the new implementations of the periodogram. During the inspection of signals to detect periodicities, we also estimate the PSD.

\begin{figure*}[htp]
    \centering
    \includegraphics[width=18cm]{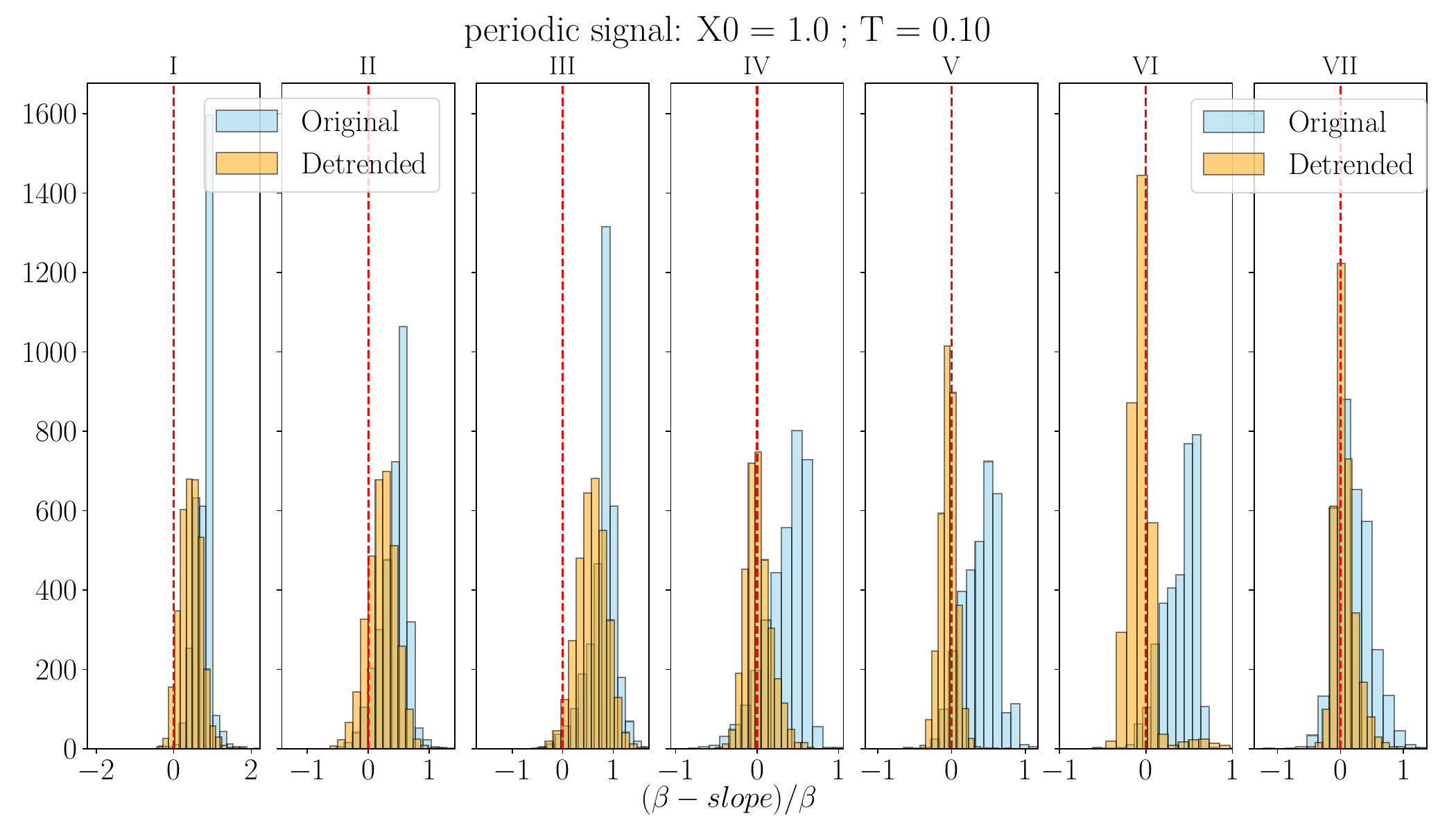}
    \caption{Statistics of the PSD estimates obtained from the periodogram of a sample of simulated signals, after injecting a periodic signal of amplitude of the standard deviation of the original signal, and period of 0.1 times the signal duration, using the classic (blue) and new (orange) implementations for each sampling tested.}
    \label{fig:PSDperiodicity}
\end{figure*}

Therefore, we can examine the effect of injecting a periodic signal into our simulated signals on the PSD estimate. Figure \ref{fig:PSDperiodicity} presents histograms showing the discrepancies between the PSD of the simulated signals and the PSD estimates. We observe a slight increase in the PSD discrepancy due to the effect of periodicity on the periodogram distribution, while maintaining consistent results with the analysis conducted without periodicity, as shown in Figure \ref{fig:PSD_Fitting}.

\section{On the Automatic Identification of the frequencies affected by noise.}
\label{Ap:NoiseFreq}

Our testing pipeline generates a statistically significant sample of simulated signals to assess the performance of the periodogram in both its classic and new implementations. However, when examining the periodogram values in the frequency domain, the noise contribution becomes dominant over the signal at higher frequencies. Therefore, it is crucial to identify these frequencies when estimating the signal PSD using the periodogram evaluated at different frequencies in the signal's frequency domain.

\begin{figure*}[htp]
\centering
    \includegraphics[width=18cm]{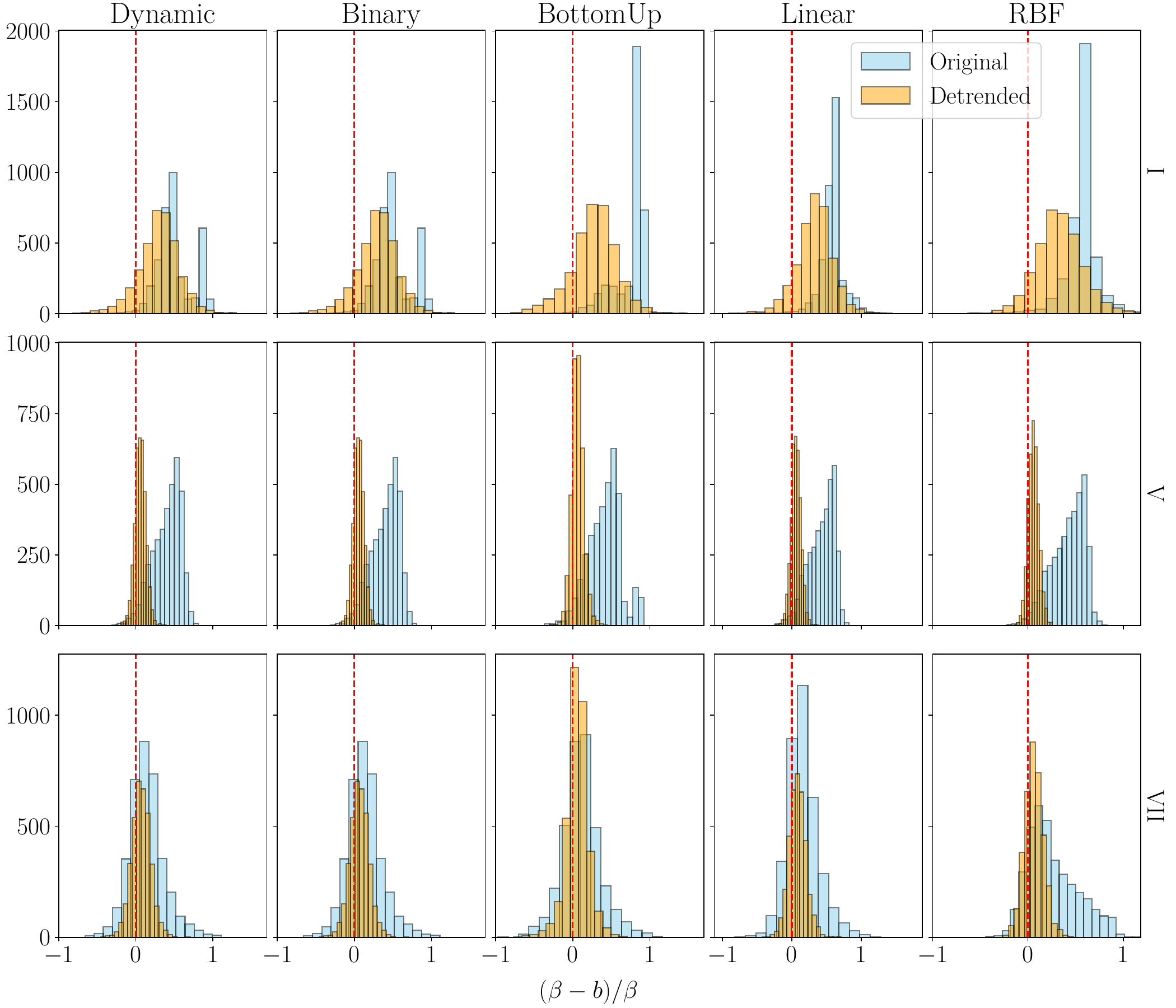}
     \caption{Statistics of the PSD estimates obtained from the periodogram of a sample of simulated signals using the classic (blue) and new (orange) implementations, for three samplings: I (top), V (middle), and VII (bottom). Five segmentation methods (from left to right: dynamic, binary, bottom-up, and change-point detection with linear and RBF kernels) are tested to identify and discard the frequencies at which the noise contribution dominates over the signal for PSD estimation.}
     \label{fig:PSD_diffmethods}
\end{figure*}

Due to the large sample of simulations, a Bottom-up signal segmentation algorithm has been used throughout the paper to identify the frequencies at which the noise contribution dominates over the signal in the periodogram. If a different signal segmentation method is used to identify these frequencies, the statistics of the discrepancies between the PSD of the simulated signals and the estimates from the periodograms differ slightly, as shown in Figure \ref{fig:PSD_diffmethods}.

In this figure, we present histograms showing the discrepancies between the PSD of the simulated signals and the PSD estimates obtained after filtering the periodogram's noise-affected frequencies using five different segmentation methods\footnote{We explore five well documented segmentation methods: dynamic (see \citet{Barry1996}), binary \citep[see][]{Bai1997, Fryzlewicz2014}, bottom-up \citep[see][]{Keogh2001, Fryzlewicz2007}, and change-point detection with linear and RBF kernels \citep[see][]{Killick2012, Celisse2018, Arlot2019}.}. This demonstrates that different methods may be more effective in identifying the noise-affected frequencies for each simulated signal, with different PSD values and samplings. Nevertheless, the new periodogram implementation consistently provides better PSD estimates compared to the classic periodogram.

\bibliography{Bibliography}{}
\bibliographystyle{aasjournal}


\end{document}